\documentclass[nologo,11pt,a4paper]{ETHpaper}

\usepackage{tabularx}                  \usepackage{MnSymbol}
\usepackage{booktabs}
\usepackage{subcaption}
\usepackage{hhline}
\usepackage[table,dvipsnames]{xcolor}
\usepackage{graphicx}                  \usepackage[english]{babel}            \usepackage[numbers,sort&compress]{natbib} 
\usepackage{setspace}
\renewcommand{\epsilon}{\varepsilon}
\usepackage{adjustbox}

\title{The role of network embeddedness \\ on  the selection of collaboration partners: \\[1ex] An agent-based model with empirical validation
}
  \titlealternative{The role of network embeddedness on the selection of collaboration partners: An agent-based model with empirical validation}
\author{Frank Schweitzer, Antonios Garas, Mario V. Tomasello, Giacomo Vaccario, Luca Verginer}
\authoralternative{F. Schweitzer, A. Garas, M. V. Tomasello, G. Vaccario, L. Verginer}
\address{Chair of Systems Design, ETH Zurich, Switzerland\\
  \url{www.sg.ethz.ch}}
\www{\url{http://www.sg.ethz.ch}}

\reference{\em (Submitted for publication)} 
\makeframing

\newcommand{\mean}[1]{\left\langle #1 \right\rangle}

\begin{document}

\maketitle

\begin{abstract}
  We use a data-driven agent-based model to study the core-periphery structure of two collaboration networks, R\&D alliances between firms and co-authorship relations between scientists. 
    To characterize the network embeddedness of agents, we introduce a coreness value, obtained from a weighted $k$-core decomposition.
  We study the change of these coreness values when collaborations with newcomers or established agents are formed. 
  Our agent-based model is able to reproduce the empirical coreness differences of collaboration partners and to explain why we observe a change in partner selection for agents with high network embeddedness.
  \end{abstract}

\section{Introduction}
\label{sec:introduction}

Collaboration is a pervasive phenomenon in the animated world.
We find it at various organismic levels \citep{ebeling2003,couzin2003self}, ranging from cancer cells to bacteria, from gregarious insects to bats and fish  \citep{jeanson2007conspecific,Schweitzer.Lao.ea1997Activerandomwalkers, deisboeck2009collective,BenJacob2015,erban2004individual,Kerth2011}.
We observe collaboration also in humans and even between human generated higher-level structures, e.g. between economic firms \citep{goyal2001r,Tomasello2013} or political parties \citep{otjes2017collaboration}.
In abstract terms, a collaborative effort most often leads to better results than the additive outcome of isolated efforts.
In economics,  phenomena such as the division of labor or the establishment of global supply chains are based on this rationale \citep{suzuki2012analyzing,takeishi2002knowledge}.
In social systems, we find for instance that publications are written by a larger group of co-authors or strategic alliances between political actors are formed \citep{Zingg2020,Pfitzner2014,van2010strategic}. 

Complexity science addresses the question whether we can detect overarching principles to characterize collaborative systems and, subsequently, mechanisms to establish collaboration.
The challenge comes with the abstraction: Instead of focusing on the specificity that distinguishes collaborating firms from, e.g., collaborating scientists, complexity science is primarily interested in the commonalities, in the general principles that constitute collaborative systems.
This aspiration entails two methodological problems:
(i) We have to define quantitative measures that allow to characterize and to compare collaboration structures across systems.
(ii) We have to identify dynamic principles that generate collaboration irrespective of the entities, or agents in general, that form the system. 

As one successful approach to problem (i), the network representation has been established.
Agents as the system elements are represented by nodes and interactions between agents as links.
The network structure then allows to define topological measures to characterize the network position of agents and to relate it to their collaboration effort.
From this perspective, successful collaboration networks should display similar features and collaborative agents could be identified by their network position. 

Such an approach would merely state a relation between observed collaborations and certain network features.
It does not explain why agents collaborate and how they establish their collaborations.
To address problem (ii), agent-based modeling has been proposed.
While this approach is deeply rooted in economics, as well as in computer science, there is no general way of developing agent-based models.
We can distinguish at least two different directions \citep{ACS-econ-netw-2009}.
The first one starts from the economic perspective \citep{jackson1996strategic,bala2000noncooperative,koenigTheorEcon2014,koenigGEB2012,koeningGEBO2011}.
Agents collaborate because they obtain a benefit.
This requires to define utility functions, i.e. costs and benefits for agents.
Further, one has to define how agents evaluate current and expected outcomes, what information they take into account and how they make decisions.
This is mostly done in a formal manner that allows to analyze mathematical properties of the model, albeit with restrictions for the chosen mathematical expressions.   
Such an approach can replicate certain topological features of observed collaboration networks, which lends some evidence to the underlying assumptions about utilities and decision rules.
At the same time, all results crucially depend on these assumptions.
Thus, instead of obtaining a general picture of how collaboration structures evolve, we mostly learn what distinguishes the utilities and decisions of firms from, e.g., scientists.

Therefore, in the second approach to agent-based modeling agents have a set of possible rules, which they follow with a certain probability \citep{ACS2016,Perra2014}.
This set of rules is neither complete, nor exclusive.
It is rather motivated by empirical observations of possible actions that agents can choose in a given situation.
Importantly, the probabilities to follow certain rules are obtained from data.
We therefore call this a data-driven modeling approach. 
The rules are in some sense ``universal'', i.e. they apply to different collaborative systems, while the probabilities reflect the specifics of the system. 

Similar to the first perspective, the validity of the modeling approach is measured against its ability to reproduce real-world collaboration structures.  
Succinctly, if problem (i) is solved by means of a network representation that gives us structural measures to compare different collaboration systems, then solving problem (ii) by means of  data-driven agent-based modeling allows to compare mechanisms to establish collaborations across systems.

In our paper, we illustrate the power of our approach by modeling and analyzing two very different collaboration systems: R\&D (research and development) collaborations between firms and co-authorship relations between scientists.
To show that the same quantitative characterization and the same dynamic assumptions to form collaborations can be applied to systems from different domains, we use an agent-based model, i.e. firms and scientists are abstracted as agents in the following.
The details of our agent-based model and its calibration for the two different systems are explained in Section~\ref{SI-SEC:Model}.  

In this paper, we focus on one specific question, namely how the network position of agents affects the selection of collaboration partners. 
Our study is motivated by the empirical observation that collaboration networks show a pronounced core-periphery structure, where a small, but highly integrated core of agents coexists with a large and sparse periphery.
To quantify the network embeddedness of agents, in Section \ref{sec:network-analysis}, we introduce the coreness value to measure the distance from the core.
We then study how differences in coreness values evolve if agents start new collaborations. 

From a dynamic perspective agents entering the network usually do not start from the core, but from the periphery.
That means, during the evolution of the network some agents manage to better integrate themselves into the network, but others not.
This brings up an important question: do agents follow specific strategies to improve their network embedding?
If they do so, then in an agent-based model we should find that their actions could not be captured by the simple probabilistic rules we apply for the ``normal'' agents.
If, on the other hand, they do \emph{not} follow specific strategies to enter the core, then we could argue that their better network integration is the result of \emph{chance} more than of strategic choice. 
Our detailed discussion in Sections~\ref{sec:impr-netw-embedd}, \ref{sec:discussion} shows that our agent-based model is able to reveal the feedback mechanisms that lead to the observed practice in partner selection. 

The remainder of the paper is divided as follows.
In Section~\ref{SI-SEC:Model}, we present the agent-based model together with an overview of the collaboration data used to calibrate the model parameters.
In Section~\ref{sec:network-analysis}, we investigate the empirical network embemdeness of firms and of scientists and compare it with the outcome of our agent-based model.
This is followed by a discussion of the results in Sections~\ref{sec:impr-netw-embedd}, \ref{sec:discussion}.

\section{Modeling the formation of collaboration networks}
 \label{SI-SEC:Model}

\subsection{Agent-based model of collaboration networks}
\label{sec:agent-based-model}

In the following, we utilize a recently proposed agent-based model which was already applied to collaborations between firms \citep{tomasello2014therole,tomasello2014therole} and between scientists \citep{Tomasello2017}.
We consider a multi-agent system with $N$ agents.
They represent either firms in an R\&D network or scientists in a co-authorship network, and links between agents represent collaborations.
Collaborating agents form groups of various size $m$, i.e. R\&D alliances or co-authorship teams,  which appear as fully connected cliques in the collaboration network.  

Our model uses two macroscopic features of empirical collaborations as input, the distribution of agents'  activities, $P(a)$ (see Section~\ref{sec:calibr-agent-based}), and the distribution of alliance sizes, $P(m)$. 
Activity defines the propensity of each agent to initiate a collaboration.
From the distribution $P(a)$, we initially sample without replacement an \emph{activity} value $a^{i}$ for each agent.
During the simulations at every time step agent $i$ initiates a collaboration with probability $p^{i}\propto a^{i} \mathrm{d} t$.
Thus, at each time step the number of active agents is $N_{A} \propto \langle a \rangle N \mathrm{d} t$, where $\langle a \rangle$ is the average agent activity.
Upon activation, an agent becomes an \emph{initiator}, i.e. selects the number of partners, $m$, with whom the collaboration is formed. 
This value of $m$ is sampled without replacement from the empirical distribution of collaboration sizes, $P(m)$. 
The second fundamental attribute of agents is their \textit{label} $l^{i}$.
The \emph{label} attribute is used to model the participation of an agent in different groups with shared practices and/or behaviors.  For the case of firms forming R\&D alliances, labels translate to membership, ``clubs'' or ``circles of influence''.
For co-authorship teams, labels indicate specific scientific specializations. 

Labels do not change over time, but can propagate to other agents. 
We assume that collaborations allow the transfer of labels to those agents that are not labeled yet.
Specifically, at the beginning of a simulation, all agents are \textit{non-labeled}, i.e. they are \emph{newcomers} with a blank membership attribute.
Once they received a label, we denote them as established agents, or \emph{incumbents}. 
A newcomer can obtain its label in two ways: (i) the agent either receives the label from another agent, if the latter initiates an collaboration (label propagation), or (ii) it takes an arbitrary and unique label when it becomes active for the first time (label generation).

This label propagation process is mapped to the formation of collaborations by means of five probabilities for link creation. 
If the initiator of a collaboration, chosen by its activity, is a newcomer (non-labeled), it links to a labeled agent with probability $p^{N\!L}_{l}$, or to another non-labeled agent with probability $p^{N\!L}_{nl}$.
If the initiator is an established agent (labeled), it has three options to form a link.
It can (i) link to an agent with the \emph{same} label with probability $p^{L}_{s}$, (ii) link to an agent with a \emph{different} label with probability $p^{L}_{d}$, or (iii) link to an agent without a label with probability $p^{L}_{n}$. 

Because the number of collaboration partners, $m$, is already given from the sampling, the above five probabilities decide how many of the $m$ partners come from each of the three \emph{partner categories}: \emph{same (s)/ different (d)/ no label (nl)}.
But the link probabilities alone are not sufficient to reproduce the features of empirical collaboration networks.
We need an additional dynamic rule to actually \emph{select} the partners \emph{within} the three partner categories.
Here we use a linear preferential attachment rule, where the probability to attach to a node $j$ linearly scales with its degree $d^{j}$, i.e. $\Pi(d^{j}) \propto d^{j}$. 
This preferential selection affects only incumbents that are already assigned to a partner category, as by definition newcomers are non-labeled and have no previous partners ($d^{j}=0$). 
Therefore, if the initiator connects to a newcomer, the partner is selected among all non-labeled nodes with equal probability. 

Because the preferential attachment rule applies to agents with very different activities, this results in a reinforcement dynamics which is important to understand the emergence of the core-periphery structure of the network.  
If agents have a high activity, they also have more collaborations over time, and therefore a higher (weighted) degree in the collaboration network.
The linear preferential attachment rule implies that within the partner categories ``\emph{same/different} label'' agents with a higher degree are also chosen with higher probability.
This further increases their degree or at least the weight of the link in case of repeated collaborations, which eventually improves the network embeddedness of these agents.

When the partner selection process is complete, all $m$ partners are mutually connected, forming a fully connected clique of size $m+1$. 
This reflects the meaning of R\&D consortia or of co-authorship teams. 

To summarize, 
our agent-based model is an \emph{activity driven model}, i.e. from the empirical distribution of activities agents get assigned a (fixed) activity $a^{i}$ to form collaborations. 
Obviously, in a stochastic simulation agents with a higher activity are on average chosen \emph{earlier} and \emph{more often}.
This generates a first mover advantage because such agents can increase their degree, i.e. the number of collaborations, early on.
In the beginning, they also get a higher chance to propagate their \emph{label} to other (unlabeled) agents. 

We note that our agent-based model does not make strategic assumptions about collaborations.
Instead, the decision of agents in establishing links with newcomers or incumbents are modeled only by means of the mentioned five probabilities, which need to be calibrated for firms and for scientists, separately. 
Therefore, this is an ideal null model to test whether the 
observed dynamics of the resulting collaboration network needs strategic agent considerations as an explanation.

\subsection{Calibration of the agent-based model}
\label{sec:calibr-agent-based}

In order to calibrate the mentioned probabilities for link formation, we need to use different data sets about collaborations of firms and of scientists.
This calibration procedure was carried out and described in detail in previous publications, therefore we only summarize it here.

\paragraph{Data sets. \ }
For firm collaborations we use Thomson Reuters' {\it SDC  Platinum alliances database}.
This contains all publicly announced R\&D partnerships (``alliances'') between 1984 and 2009 with a resolution of 1 year. 
In total we have $E=14,829$ alliance reports, referred to as collaboration events $E$, involving $N=14,561$ firms.

For the co-authorship collaborations of scientists, we
use the data set from the American Physical Society (APS) about papers published in any
APS journal, namely Physical Review Letters, Reviews of Modern Physics, and all
Physical Review journals \citep{}.
This data set is quite large, it spans 110 years (1895-2004) and contains 
$N=226,724$ unique authors and $E=1,567.084$ publications, i.e. collaboration events. 
For the empirical study of network embeddedness we used the full data set.
But for the calibration of the agent-based model we only selected papers published between  1984 and 2009 in specific research areas identified by their PACS code.
We have restricted ourselves such that the time periods for both data sets are the same and  the collaboration networks are of comparable size.
In Table~\ref{table:collaboration_networks_sizes}, we present the example of PACS 
42 (Optics), for which we have 
in total $E=20,105$ publications involving $N=27,436$ scientists.

But for the calibration of the agent-based model we only selected papers published between 1984 and 2009 in specific research areas identified by their PACS code.

\paragraph{Distributions. \ }

From these data sets, we calculate the distribution of collaboration sizes, $P(m)$, as well as the activity distribution, $P(a)$, both for the firms and for the scientists.
These distributions, which are used as an input for the agent-based simulations, are very broad \citep{Tomasello2017}. 
In particular, the activity distributions span several orders of magnitude.
Here the empirical activity of a given agent $i$ at time $t$ is the number of collaboration events, $e_{i,t}^{\Delta t}$,  involving agent $i$ during a time window $\Delta t$ (in years) ending at time $t$ divided by the total number of
collaboration events, $E_{t}^{\Delta t}$, involving any agent during the same period of time.
We mention that the activity distributions are very stable regardless of the chosen $\Delta t$.

\paragraph{Network reconstruction. \ }

In a next step, we reconstruct the aggregated collaboration networks for firms and for scientists.
We emphasize that these are undirected, but \emph{weighted} networks, where the weight $w^{ij}$ gives the number of collaborations between agents $i$ and $j$ over the whole time.
Further, we note that the number of links, $L$, is not the same as the number of collaboration events, $E$.
A publication co-authored by 4 scientists, for instance, would count as one collaboration event of size $m+1=4$, but it generates 6 links between the 4 involved scientists.

\begin{figure}[htbp]
\begin{center}
(a)\includegraphics[width=0.40\textwidth]{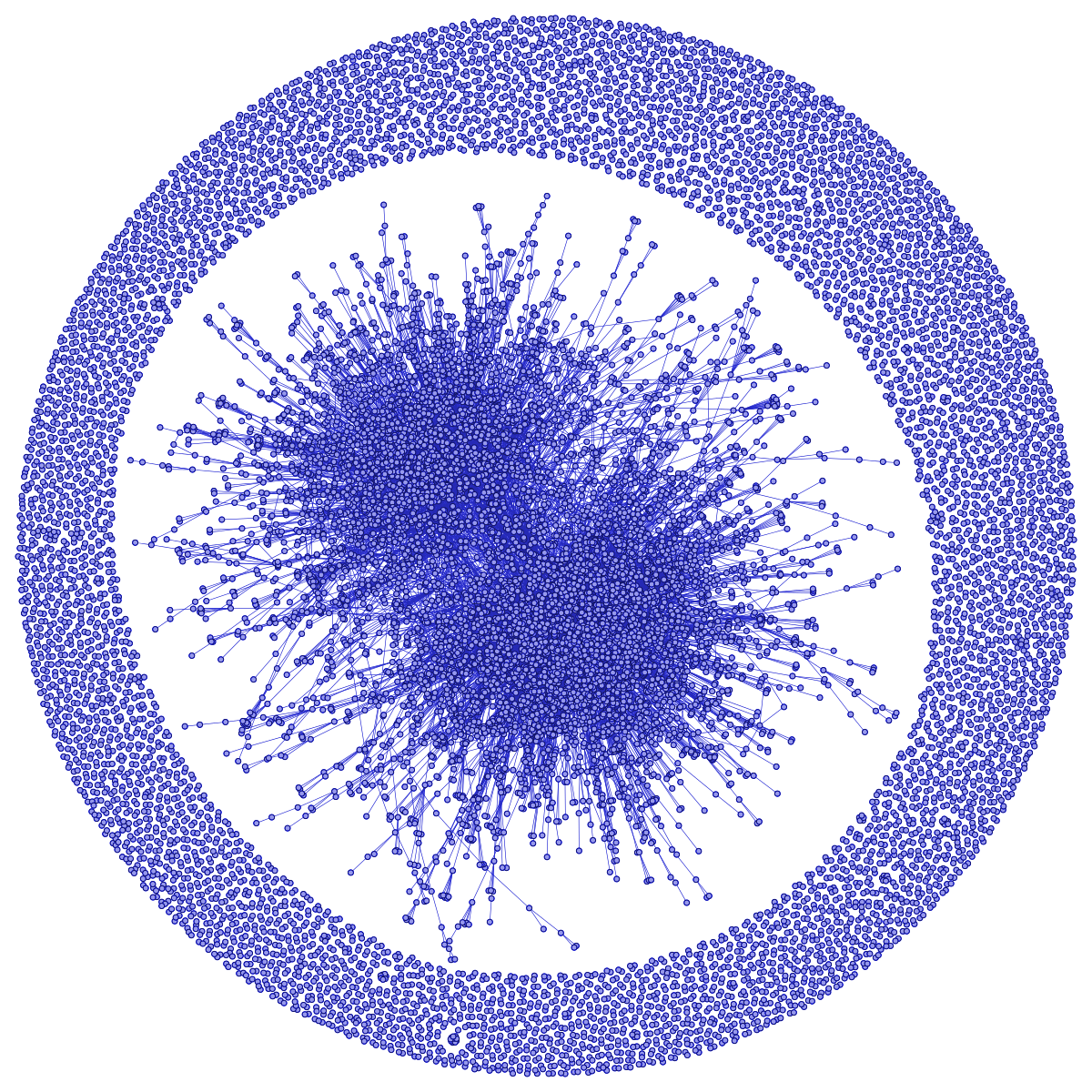}
(b)\includegraphics[width=0.40\textwidth]{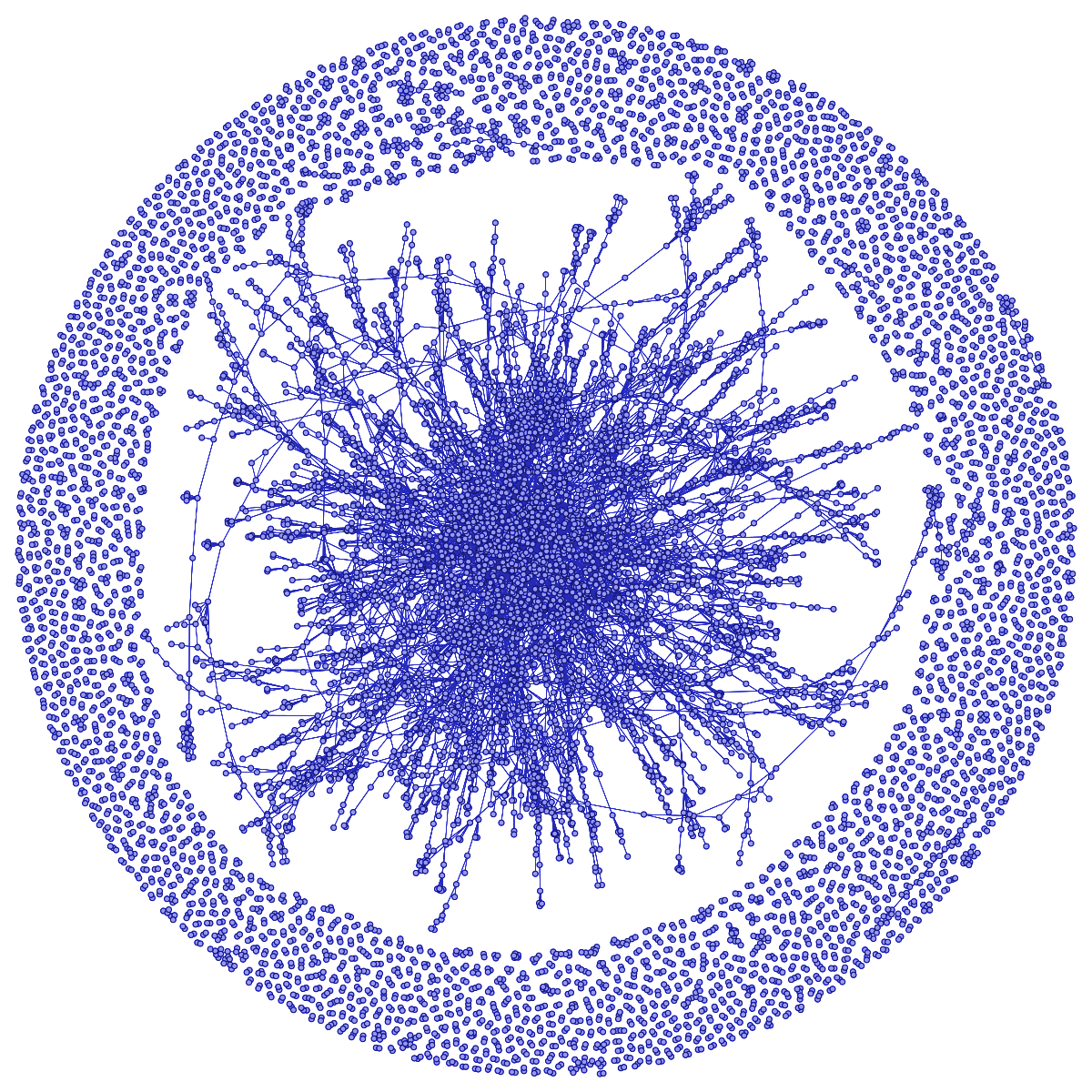} 
\end{center}
\vspace{-4pt}
\caption[]{Illustration of the collaboration networks of firms (a) and of scientists (b). Data: (a) complete R\&D network with about 14,000 nodes and 21,000 links, (b)  co-authorship network sampled from the full data set with about 11,000 nodes and 32,000 links.
  }
\label{fig:network_plots}
\end{figure}

Figure \ref{fig:network_plots} shows these two networks as unweighted networks.
There are two important observations: (i) Both collaboration networks have a largest connected component (LCC) and a large number of small disconnected clusters.
(ii) The LCC itself shows a prominent \emph{core-periphery} structure, which may be difficult to see because the LCC is quite dense.
But for the R\&D collaborations, for instance, we note that the inner core contains less than 250 out of 14.000 firms. 
When we analyze the evolution of network positions of agents in the following, these obviously refer to the core-periphery structure of the LCC, only.

\paragraph{Calculating average quantities. \ }

From the observed aggregated networks, we calculate three mean quantities: (i) Average degree $\mean{d}^{o}=2N/L$, where $N$ is the total number of nodes and $L$ is the total number of links.
The normalization factor of 2 results from the fact that each link connects two nodes. (ii) Average path length $\mean{l}^{o}$.
A path is formally defined as a sequence of nodes, where any pair of consecutive nodes is connected by a link, i.e. loosely speaking the path length is the number of steps to reach a node over the network from a given starting point. 
(iii) Average clustering coefficient $\mean{c}^{o}$.
The local clustering coefficient of a node captures the fraction of its neighbors that are directly connected, i.e. loosely speaking it counts the fraction of triangles in a neighborhood.
$\mean{c}^{o}$ is the mean of all local clustering coefficients. 

From the R\&D network, we further obtain the degree distributions $P^{o}(d)$, which give the number of collaboration links of firms, to later compare it with our simulation results in Section~\ref{sec:results-comp-simul}. 

\paragraph{Calculation of link probabilities.\ }

To determine the quantities  $p^{N\!L}_{l}$,
$p^{N\!L}_{nl}$, $p^{L}_{d}$, $p^{L}_{s}$, $p^{L}_{n}$, we run agent-based simulations with all possible combinations of values. 
We take $N$ and $E$ as input, further we sample agent activities $a^{i}$ and collaboration sizes $m$ from the respective distributions.
From each simulation, we construct the respective network on which we calculate the mean values $\mean{d}^{s}$, $\mean{l}^{s}$, $\mean{c}^{s}$.
We can then determine the error $\epsilon$ from the differences between the observed and the simulated quantities.
We require that these errors have to be smaller than a threshold $\epsilon^{0}$.
For all probability combinations we perform 25 simulations and select the combination that gives us the highest fraction of networks that match the criterion $\epsilon<\epsilon^{0}$.

\paragraph{Optimal simulation values. \ }

From our calibration procedure, we have determined for each data set those link probabilities (indicated by $^{*}$) that would generate an \emph{optimal} network in the sense that the expected error between observations and simulations is minimized.
We emphasize that this only refers to the \emph{averaged} quantities defined above.
We have only  used \emph{global information} for the calibration, no specific knowledge about link preferences, etc.

Table~\ref{table:collaboration_networks_sizes} summarizes our values obtained, together with some characteristics of the data. 

\renewcommand{\arraystretch}{1.2}
\begin{table}[htbp]
\centering
\tabcolsep=0.14cm
\begin{tabular}{| c || m{35pt}|m{35pt}|m{35pt} || m{35pt}|m{35pt}|m{35pt}|m{35pt}|m{35pt}|}
\hline

  \textbf{Collaboration} & \centering \boldmath${N}$ & \centering \boldmath${L}$ & \centering \boldmath${E}$ 
& \centering \boldmath$\quad p^{*L}_{s}$ & \centering \boldmath$p^{*L}_{d}$ & \centering \boldmath$p^{*L}_{n}$ & \centering \boldmath$p^{*N\!L}_{l}$ & \centering \boldmath$p^{*N\!L}_{nl}$ \\
\hline
firms & \centering 14,561 & \centering 21,572 & \centering 14,829 & \centering 0.30 & \centering 0.30 & \centering 0.40 & \centering 0.75 & \centering 0.25 \\ 
\hline
{scientists} & \centering 27,436 & \centering 94,961 & \centering 20,105 & \centering 0.60 & \centering 0.05 & \centering 0.35 & \centering 0.35 & \centering 0.65 \\ 
\hline
\end{tabular}
\vspace{2mm}
\caption{Networks for R\&D and for co-authorship collaborations (PACS 42): Number of nodes $N$, links $L$, collaboration events $E$. 
  Optimal sets of link probabilities to simulate the collaboration networks,  indicated by $^*$. 
}
\label{table:collaboration_networks_sizes}
\end{table}

\paragraph{Interpretation of probabilities. \ }

We note that the values of the link probabilities allow for an interpretation \citep{Tomasello2017}.
For R\&D collaborations between firms, for instance, we found 
that incumbents follow a balanced collaboration strategy.
30\% of their collaborations are with agents in the same circle of influence ($p^{*L}_{s} = 0.3$), 30\% with agents in a different circle of influence ($p^{*L}_{d} = 0.3$) and 40\% with newcomers ($p^{*L}_{n} = 0.4$), represented by non-labeled agents.
At the same time, newcomers show a strong tendency to connect to incumbents ($p^{*N\!L}_{l} = 0.75$), as opposed to a low linking probability with other newcomers ($p^{*N\!L}_{nl} = 0.25$).

Comparing these values with our findings for co-authorship networks, we note both similar and different tendencies.
First, established agents prefer to form links with other established agents ($p_{s}^{*L}+p_{d}^{*L}\geq 0.6)$. 
Secondly, when forming a link with an established agent, the initiator tends to select an agent with the same label ($p_{s}^{*L}>p_{d}^{*L}$). 
This tendency is much more pronounced in co-authorship networks, i.e. to choose a co-author from a different community is less likely than to choose a firm from a different circle of influence. 
Thirdly, in co-authorship networks newcomers have a stronger tendency to link with other newcomers ($p_{nl}^{*NL}>p_{l}^{*NL}$), while for R\&D collaboration the opposite is true: newcomers preferably link to incumbents  ($p_{l}^{*NL}>p_{nl}^{*NL}$).
This difference could be explained with different entry barriers for publications and patents.
Scientists that are newcomers to the publication market can still find opportunities to  write papers together, whereas firms that are newcomers to R\&D activities may find it more difficult to generate patents. 

\paragraph{Validation criteria. \ }

We can now use the optimal values for the link probabilities to simulate the evolution of synthetic networks over time. 
The results, averaged over many runs, are then compared to findings from the empirical network.
A mere match between empirics and simulations would not be sufficient to conclude that the underlying rules of link formation are correct, because we cannot rule out that other sets of rules would lead to equally good results.
However, a good agreement lends strong evidence to our agent-based approach, in particular if the \emph{same} model is also able to replicate \emph{different} empirical findings. 

In fact, we have already demonstrated that our agent-based model can  reproduce the empirical \emph{distributions} of path lengths, clustering coefficients, degrees, component sizes, etc. in two different domains, collaborations between firms and between scientists \citep{Tomasello2017}. 
This is remarkable because for the calibration we have only used information about the \emph{mean values}, which are not sufficient to characterize any distribution.
With the simulation replication of these different findings we claim that the model indeed captures the essence of the \emph{dynamic interactions} underlying these collaboration networks, rather than simply (over)fitting free parameters to available observations.

\section{Dynamics of network embeddedness}
\label{sec:network-analysis}

\subsection{Measuring network embeddedness}
\label{sec:rules}

In this paper, we focus on a quantitative measure for network embeddedness, which we take as a benchmark here.
To characterize the topological embedding of nodes in a network, various centrality measures have been proposed that are also partially correlated \citep{Freeman1979}.
Because our networks show a clear core-periphery structure, we have introduced a  centrality measure for weighted networks, called \emph{coreness} $C_{C}^{i}$ \citep{Garas2012a}. 
Versions for unweighted networks have been used earlier in social network analysis~\cite{Seidman1983,Borgatti2000}.
In a recent paper \citep{vaccario-verginer-2022-networ}, we have shown that our coreness measure is well suited to quantify network embeddedness, in particular when compared to other centrality measures. 
It also strongly correlates with the success of agents, as quantified by non-topological measures such as the number of patents for firms or the number of citations for scientists.
Hence, we have argued that our measure of embeddedness can be used to characterize the innovation potential of firms \citep{vaccario-verginer-2022-networ}.

\begin{figure}[htbp]
  \centering
  \scalebox{0.8}{
    \includegraphics[width=4.4cm]{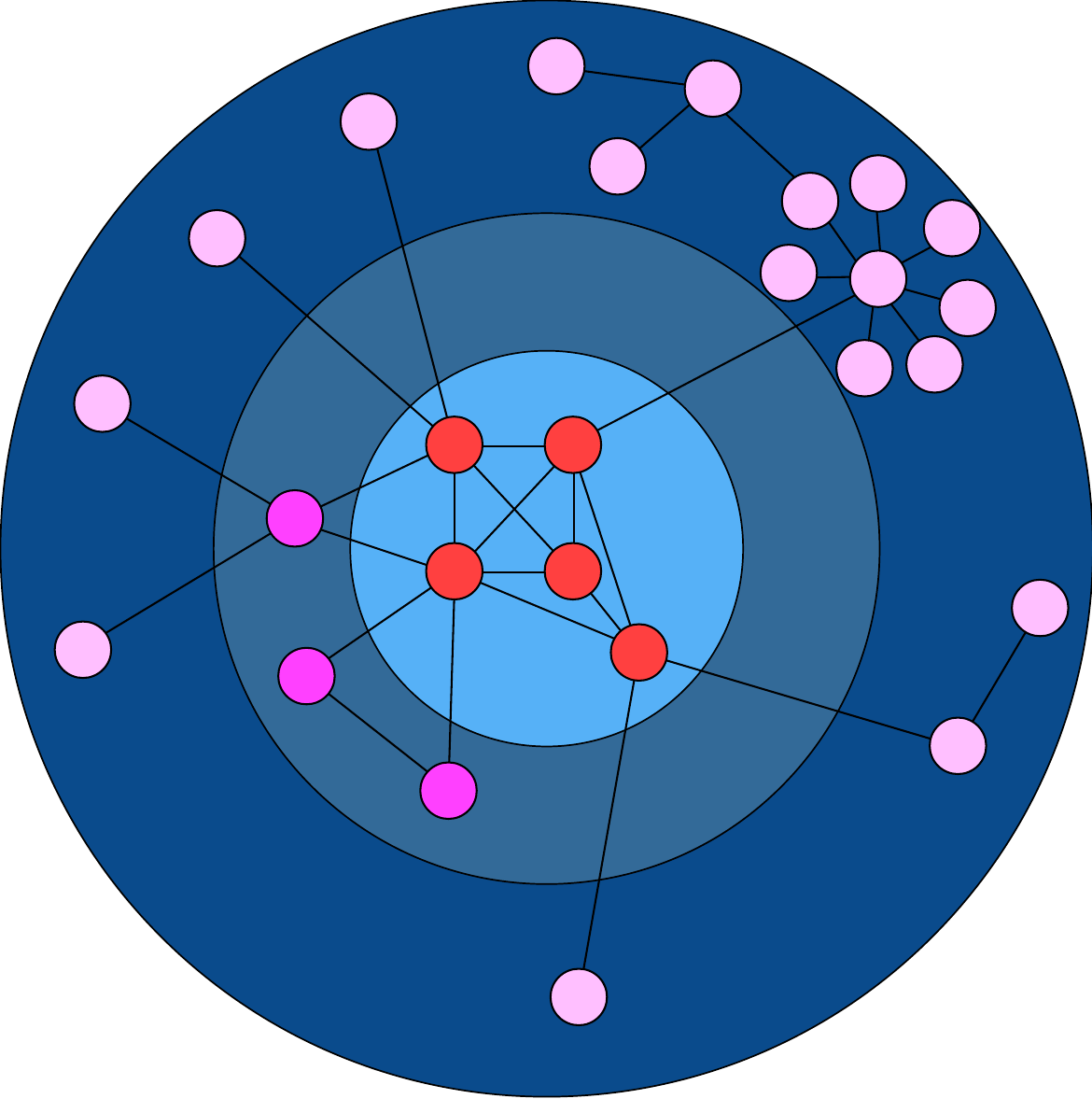}
  $\quad$\vrule$\quad$ 
  $k_{s}$=1
    \includegraphics[width=4.4cm]{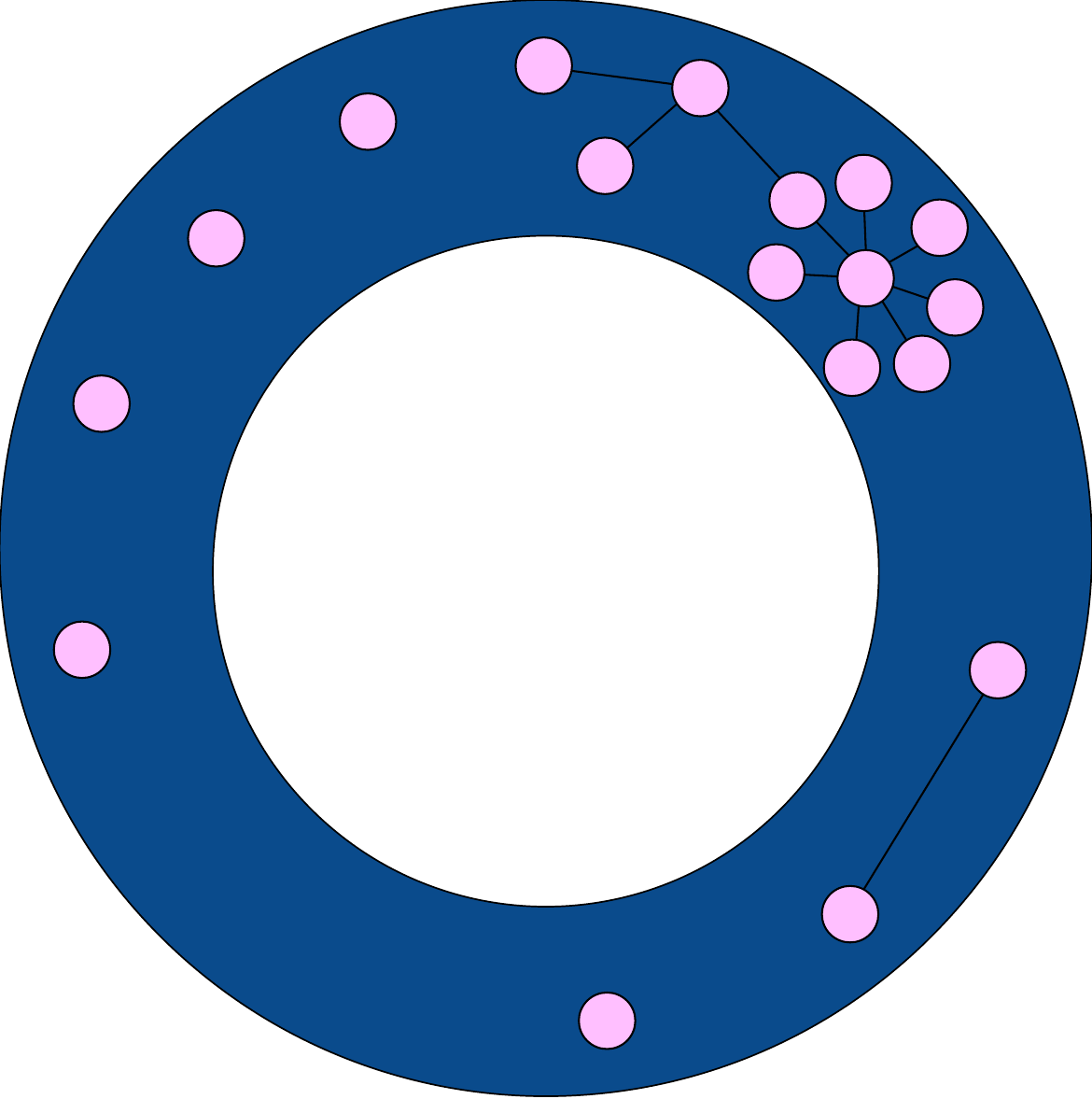}
    \adjustbox{lap={\width}{0em}}{$\quad$ $k_{s}$=2}
    \includegraphics[width=2.76cm]{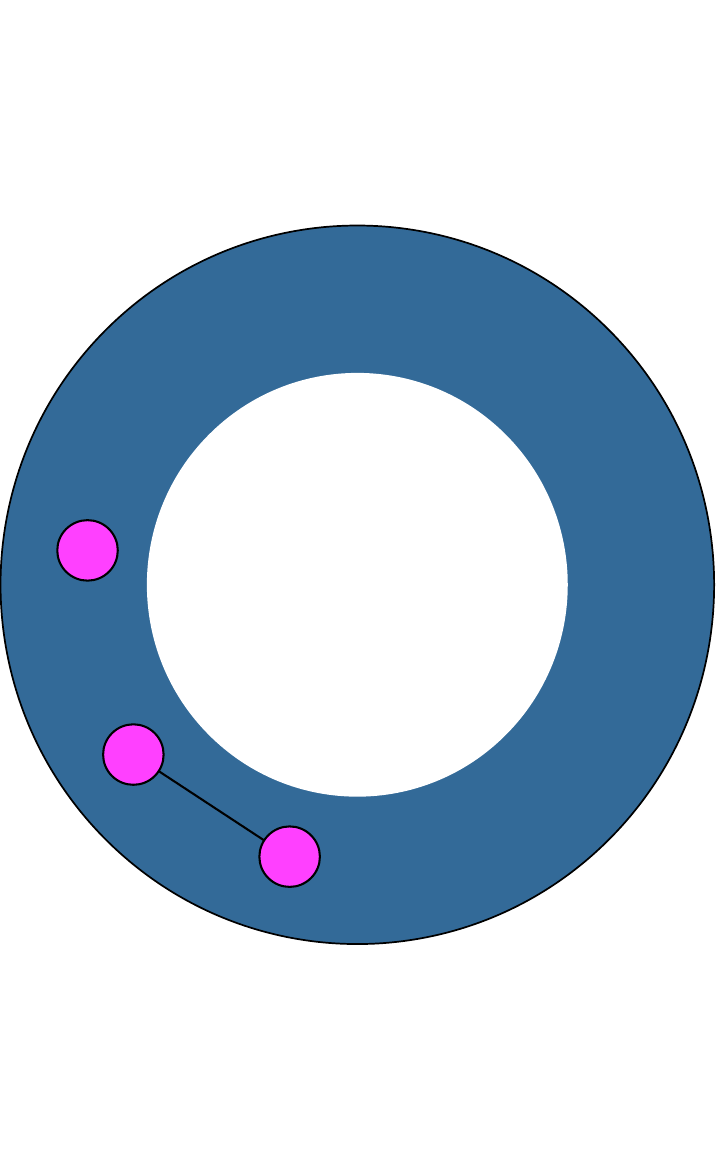}
    $\quad$ $k_{s}$=3
    \includegraphics[width=1.6cm]{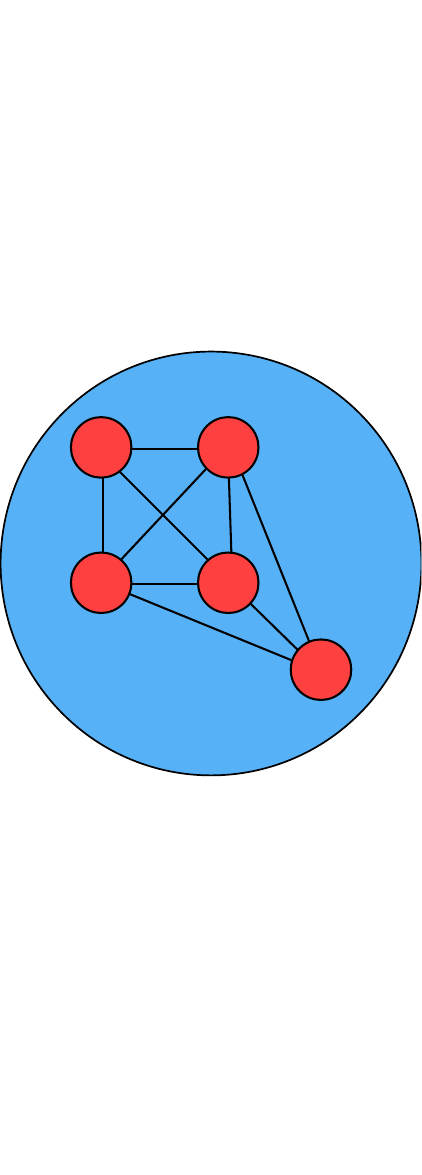}
    }
    \caption{Illustration of the weigthed $k$-core decomposition.}
  \label{fig:kcoresketch}
\end{figure}
We do not replicate the argumentation here, but only explain how coreness values for agents are obtained from the reconstructed network. 
We use the so called $k$-core decomposition (see Figure~\ref{fig:kcoresketch}), which  \emph{recursively} removes all nodes with a degree less than $d$ from the network, similar to a \emph{cascade}.
It starts with $d=1$, i.e. it removes all nodes that have only \emph{one} neighbor in the networks.
The removal may leave these neighboring nodes with one additional neighbor, hence in the second step of the cascade such nodes are also removed. 
Their removal again may leave other nodes with one remaining neighbor.
Thus, in the third step they are also removed and so forth, unless the cascade \emph{stops}.
Then, \emph{all} nodes that have been removed during this cascade are assigned a shell number $k_{s}$ equal to $d$. 

Nodes with a small $k_{s}$ obviously are not well integrated in the network, whereas nodes with the largest $k_{s}=k_{s}^{\mathrm{max}}$ are in the {\it core} of the network. 
The coreness of each node is then defined as $C_{C}^{i}=k_{s}^{\mathrm{max}}-k_{s}^{i}$, i.e., it measures the distance from the core.
\emph{Low} coreness values characterize the \emph{core}, whereas \emph{high} coreness values characterize the \emph{periphery} of the network.

Instead of the unweighted $k$-core decomposition, in this paper we apply the {\it weighted $k$-core decomposition}~\cite{Garas2012a}.
It uses the {\it weighted degree}, $d'$, defined as:
\begin{equation}
  d'^{i}=\left[\left(d^{i}\right)^{\alpha} \left( \sum\nolimits_{j}^{d^{i}}{w^{ij}}
    \right)^{\beta} \right]^{\frac{1}{\alpha+\beta}}
  \label{eq:gsh.kprime}
\end{equation}
$d^{i}$ is the degree of node $i$ and $w^{ij}$ is the weight of the link between nodes $i$ and $j$.
The summation goes over all neighbors of $i$.
The free parameters $\alpha$, $\beta$ can balance the influence of the weights $w^{ij}$.
Following \citep{vaccario-verginer-2022-networ}, we set $\alpha=1$ and $\beta=0.2$.

\subsection{Empirical dynamics of network embeddedness}
\label{sec:evol-coren-valu-1}

In the following, our focus is on the \emph{evolution} of network embeddedness, as quantified by the coreness values $C_{C}^{i}(t)$ of individual agents $i$.
These values change over time either because new collaborations are established that involve agent $i$ or because the network as a whole grows.
The latter means that new agents enter the network and form new links to incumbents, i.e. established agents,  or to other newcomers. 
As the result of network growth, new $k$-shells appear, which instantaneously affect the coreness values of all agents, even if they are not establish new collaborations.

To compare coreness values at different times, we introduce the \emph{relative} coreness  $c^{i}(t)=C_{C}^{i}(t)/C_{m}(t)$, i.e. the ratio between the current coreness $C_{C}^{i}(t)$ and the maximum coreness $C_{m}(t)$ at the same time. 
$c^{i}(t)$ can have values between 0, which indicates the  very \emph{core}, and 1, which indicates the outest \emph{periphery}.  

\begin{figure}[htbp]
  \centering
(a)  \includegraphics[width=0.45\textwidth]{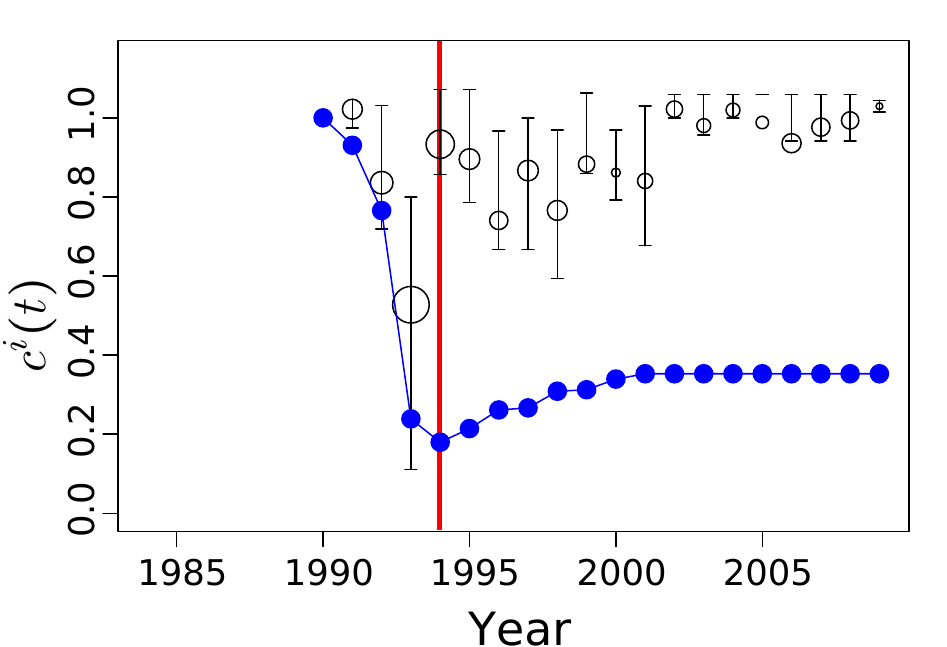}\hfill
(b)  \includegraphics[width=0.45\textwidth]{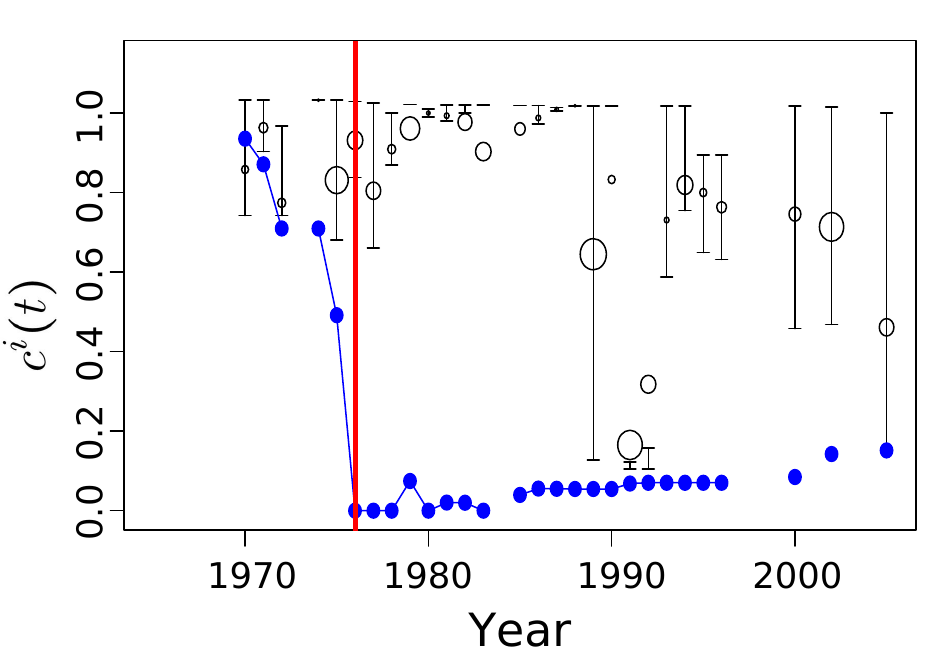}
  \caption{Evolution of relative coreness in blue for (a) Glaxo (b) Feldmann. The year in which the minimum relative coreness value is marked with a red vertical line. The average relative coreness of the partners of the two selected agents is plotted using black circles. The size of the circle is proportional to the number of partners in that year.}
  \label{fig:trajectories}
\end{figure}

Figure~\ref{fig:trajectories} shows two examples, one for a firm, the other one for a scientist, of how these relative coreness values change over time in the respective R\&D or co-authorship network. 
We see that in both cases agents start with high relative coreness values because the collaboration network still has to be established.
In the two examples, for both agents the relative coreness values then decline over time, indicating that they become part of the core as the network evolves.

We note that agents do not always keep their network position in the very core ($c^{i}=0$), as can be seen for the firm Glaxosmithcline.
It has entered the R\&D network in 1990 and reached the minimal distance to the core in 1994.
Then, it slowly moved away from the core, in particular  because \emph{other} firms managed to become better integrated into the core.
Only in 2001, the firm reached a stable network position.
Hence, for each agent we can identify a minimum relative coreness value $c^{i}(t^{i}_{c})$
corresponding to the best overall position in the collaboration network and a time  $t^{i}_{c}$ at which the maximum embeddedness of agent $i$ in the network is obtained.
This is indicated by a red line in Figure \ref{fig:trajectories}.

As an additional information, Figure \ref{fig:trajectories} also shows for the two selected agents  their relative coreness together with the relative coreness of their collaboration partners.
This allows an interesting observation.
In the early period when the two agents are rather new to the network and are thus still part of the periphery, they have a strong tendency to choose collaboration partners that have a  \emph{comparable} coreness value.
This continues until the agents reach their state of minimal coreness, $c^{i}(t^{i}_{c})$. 
For times $t>t^{i}_{c}$ we observe a change: once agents have reached the core, i.e. are well embedded in the network, they choose partners of \emph{high coreness}, i.e. newcomers or agents from the periphery.

We verify this finding by taking into account all collaborations of firms and of scientists over time. 
For each agent $i$, we calculate the relative coreness $c^{i}(t)$ for every year $t$ and the time $t^{i}_{c}$ of minimal coreness. 
We further calculate the number of collaborations with each of their partners $j$, i.e. $w^{ij}(t)$, and the total number of collaborations $A^{i}(t)=\sum_j w^{ij}(t)$ in the given year. 
Eventually, we calculate the relative coreness $c^{j}(t)$ of each of their partners $j$.
Combining all these information, we obtain the weighted average of \emph{coreness differences}:
\begin{equation}
\label{eq:2}
 \mean{d c^{i}(t)} = \frac{1}{A^{i}(t)}\sum_j {w^{ij}(t)} \left[c^{i}(t)-c^{j}(t)\right]
\end{equation}
After calculating $\mean{d c^{i}(t)}$ for all times $t$ between $t_{\mathrm{start}}$ and $t_{\mathrm{end}}$, e.g. between 1984 and 2009 for the case of firms, we divide the values according to two  time periods, \texttt{before} and \texttt{after} $t^i_c$ and average  for each of these periods separately: 
\begin{equation}
  \label{eq:3}
   \mean{d c^{i}_{\mathrm{before}}} = \frac{1}{t^{i}_{c}-t_{\mathrm{start}}}\sum_{t=t_{\mathrm{start}}}^{t^{i}_{c}} \mean{d c^{i}(t)} \;; \quad
 \mean{d c^{i}_{\mathrm{after}}} = \frac{1}{t_{\mathrm{end}}-t^{i}_{c}}\sum_{t = t^{i}_{c}}^{t_{\mathrm{end}}} \mean{d c^{i}(t)}
\end{equation}
For each agent $i$, these two values are related to the final coreness value of that agent, $C_{F}$ at $t_{\mathrm{end}}$, i.e., $\mean{d c^{i}_{\mathrm{before}}}(C_{F})$ and $\mean{d c^{i}_{\mathrm{after}}}(C_{F})$. 
Then, for each value of $C_{F}$, e.g. between 0 and 17 for the case of firms, we average the $\mean{d c^{i}}$ with the same $C_{F}$ separately \texttt{before} and \texttt{after} $t_{c}$. 

\begin{figure}[htbp]
  \centering
  (a) \includegraphics[width=0.45\textwidth]{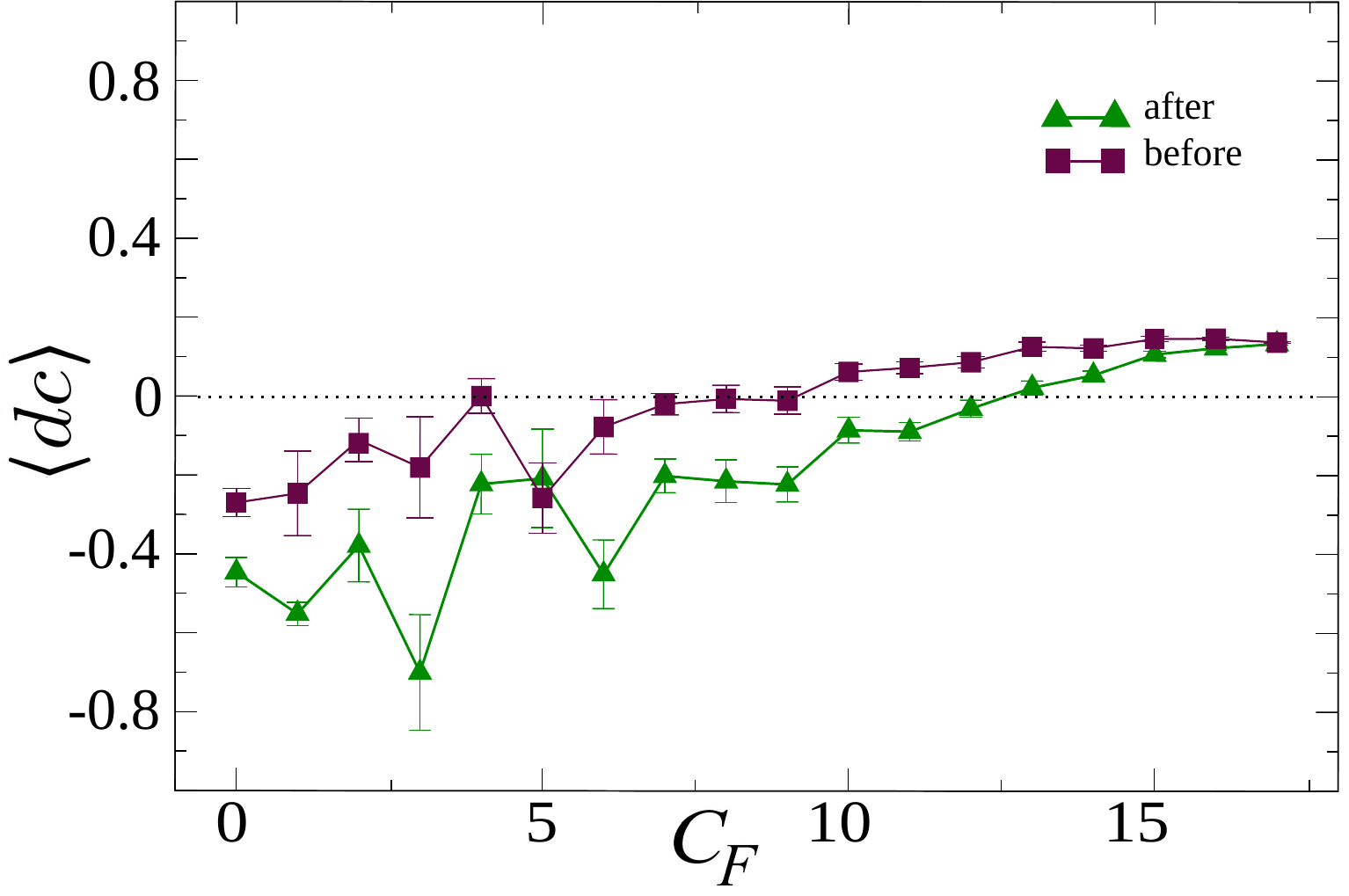}\hfill
(b)  \includegraphics[width=0.45\textwidth]{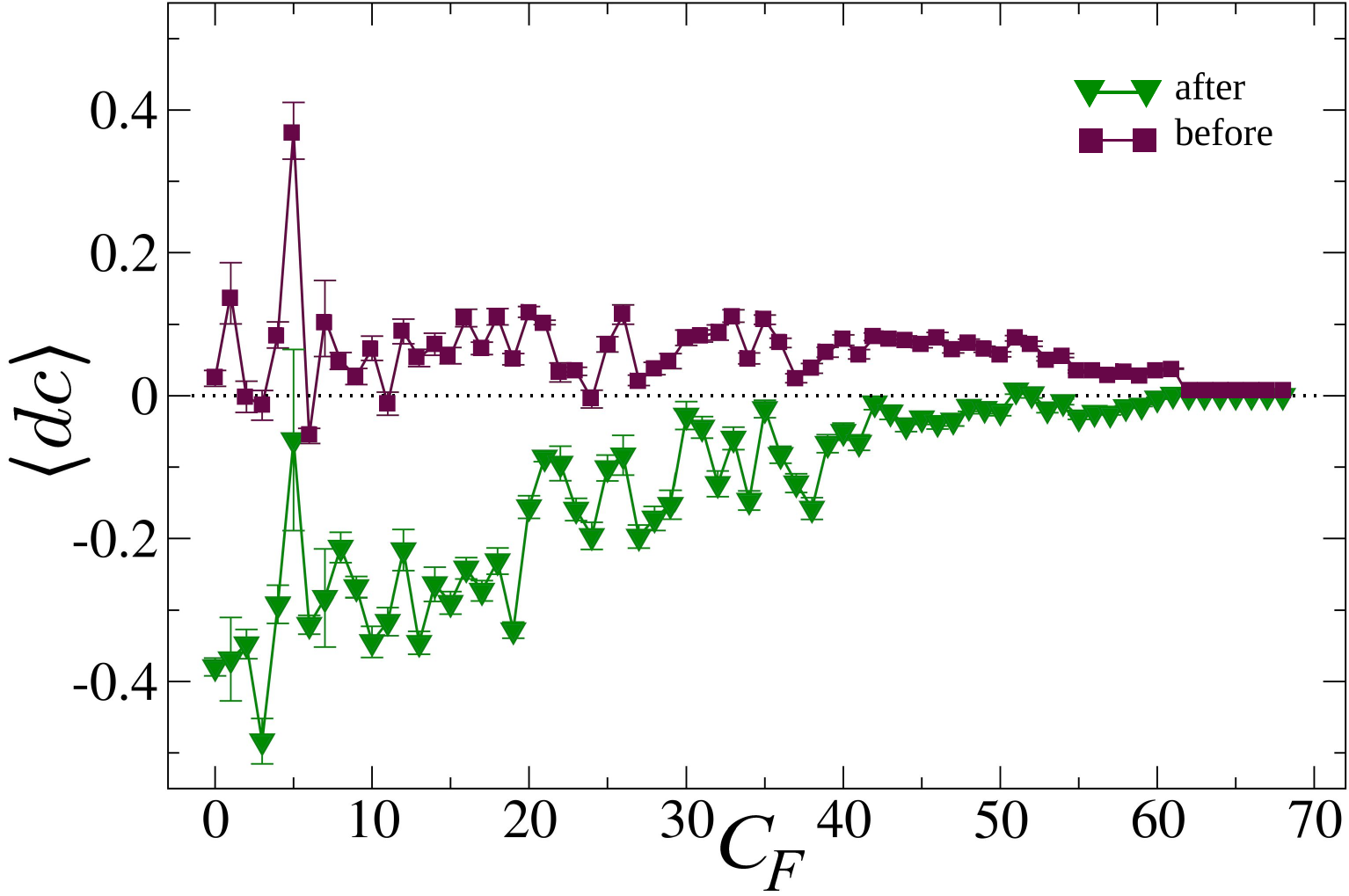}

  \caption{Average partner coreness deviation. 
  Plot of the
    average normalized partner coreness deviation $\mean{dc}$
    against $C_F$ before and after $t_c$. (a) Firms (b) Scientists.}
  \label{fig:GTS-3}
\end{figure}

The results are shown in Figure \ref{fig:GTS-3}. 
We observe that, for the two periods \texttt{before} and \texttt{after} $t_{c}$, the averaged coreness differences \emph{decrease} monotonously with final coreness $C_{F}$ and even become negative. 
Positive values mean that the initiating agent has, on average, a higher coreness than its chosen partners. 
This applies for initiators with high coreness, i.e. newcomers or agents in the periphery that strive to get a better network position by choosing better integrated partners. 
Negative values mean that this relation switches: initiating agents have on average a lower coreness, i.e. they are better integrated than their partners.
This applies for initiators with low coreness that made it to the core and confirms the previous discussion that agents closer to the core have more collaborations with newcomers or agents with high coreness. 

Looking particularly at differences between the two time periods, we find that this shift from positive to negative coreness differences becomes much stronger in the period \texttt{after} $t_{c}$, i.e., for agents that have already reached their best network position.
This means that established core agents choose even more partners with high coreness (newcomers, periphery) than agents in the period \texttt{before} $t_{c}$ that are still striving for a better network position. 

To test the robustness of this finding, we performed a random reshuffling of the collaboration links of firms while preserving the degree sequence of the empirical R\&D network. 
The dashed curves in Figure \ref{fig:GTS-3}(a) show the averaged coreness differences \texttt{before} and \texttt{after} $t_{c}$ for the reshuffled network. 
We see that the trend is the same as for the empirical network.
However, the \emph{differences} between the two curves are much larger for the \emph{empirical} network than for the reshuffled network. 
This means that the \emph{observed} change \texttt{before} and \texttt{after} $t_{c}$ is not random. 
We preformed a    two-sided Kolmogorov-Smirnov test to the distributions of the
    $\mean{dc}$ for the empirical and the reshuffled network,
    and we can reject that they are the same with $p$ = 0.056.

\subsection{Validation of the agent-based model}
\label{sec:results-comp-simul}

\begin{figure}[htbp]
  \centering
  \includegraphics[width=0.5\textwidth,angle=0]{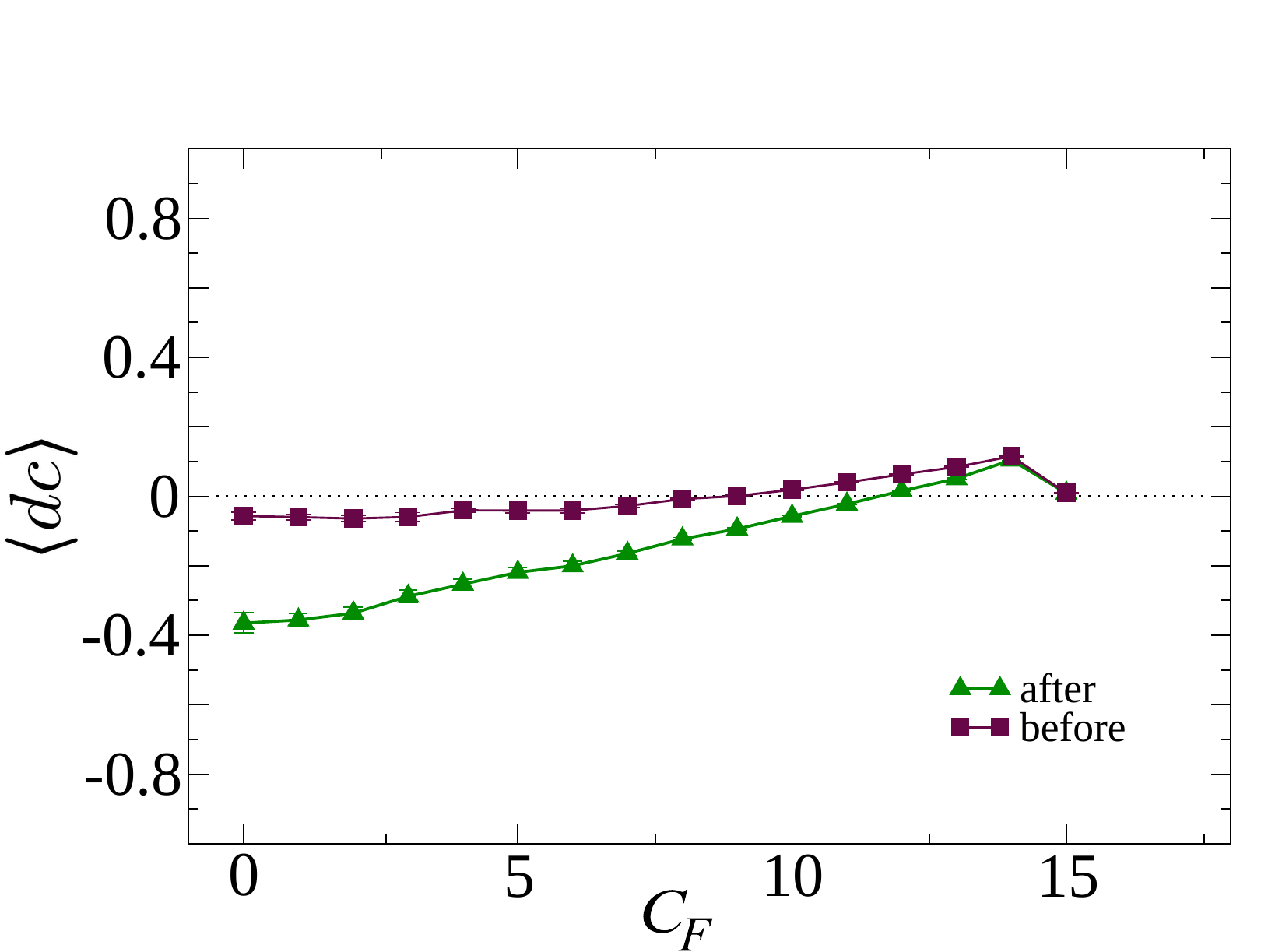}
  \caption{Average normalized partner coreness deviation $\mean{dc}$
    before and after $t_c$ obtained from the agent-based model, using the calibration from the R\&D network.}
  \label{fig:GTS-4new}
\end{figure}

To validate our agent-based model on the macro level, we have to verify that the dynamics observed in Figure~\ref{fig:GTS-3} for firms and for scientists can indeed be replicated by our model. 
Our main result is presented in Figure \ref{fig:GTS-4new} which should be compared to the empirical findings shown in Figure \ref{fig:GTS-3}. 
It demonstrates that our agent-based model generates the same pattern in partner selection that was observed empirically, namely after reaching their lowest coreness value, agents establish collaborations with more peripheral agents. 

\begin{figure}[htbp]
  \centering
(a)\includegraphics[width=0.45\textwidth]{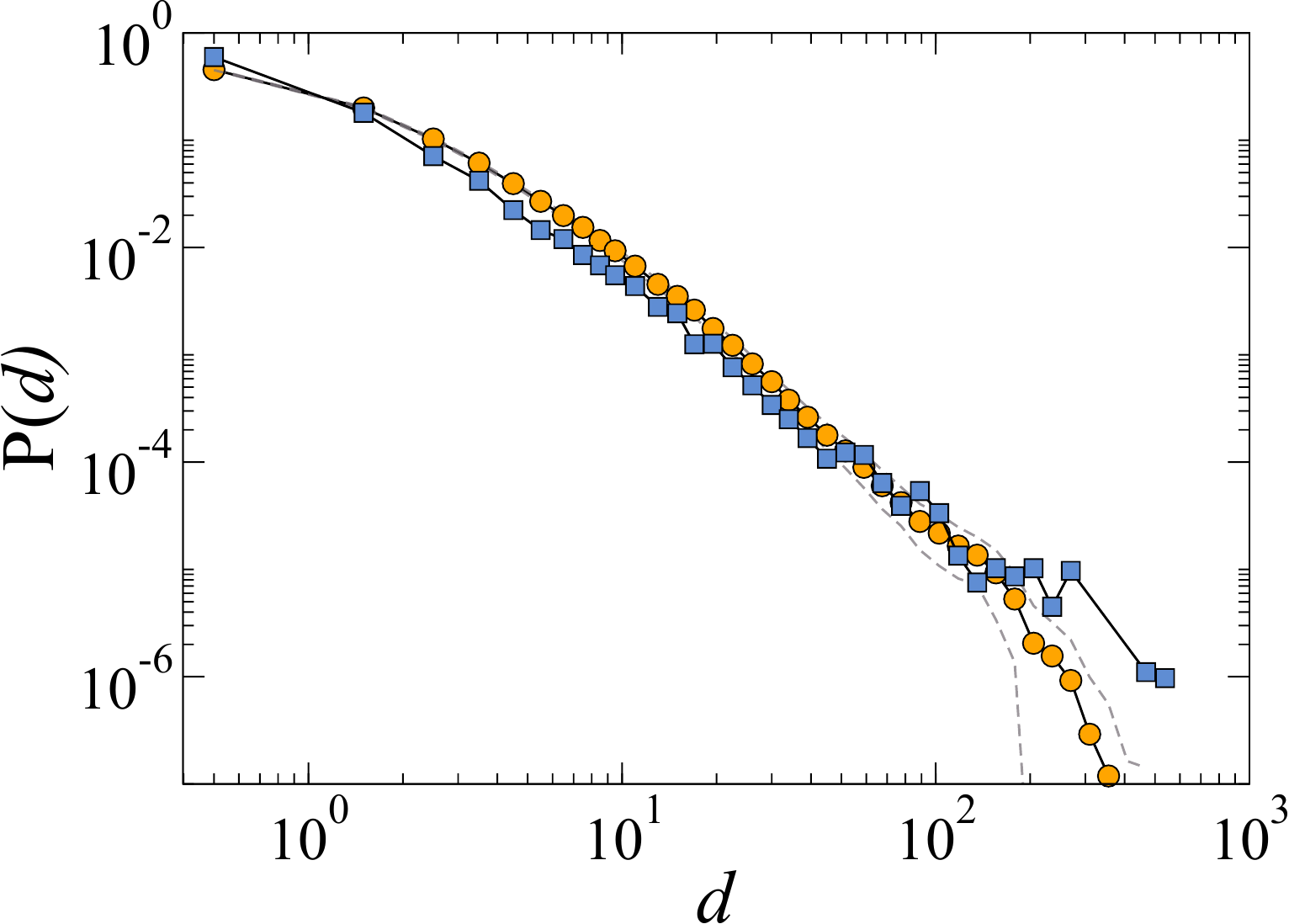}\hfil
 (b) \includegraphics[width=0.45\textwidth]{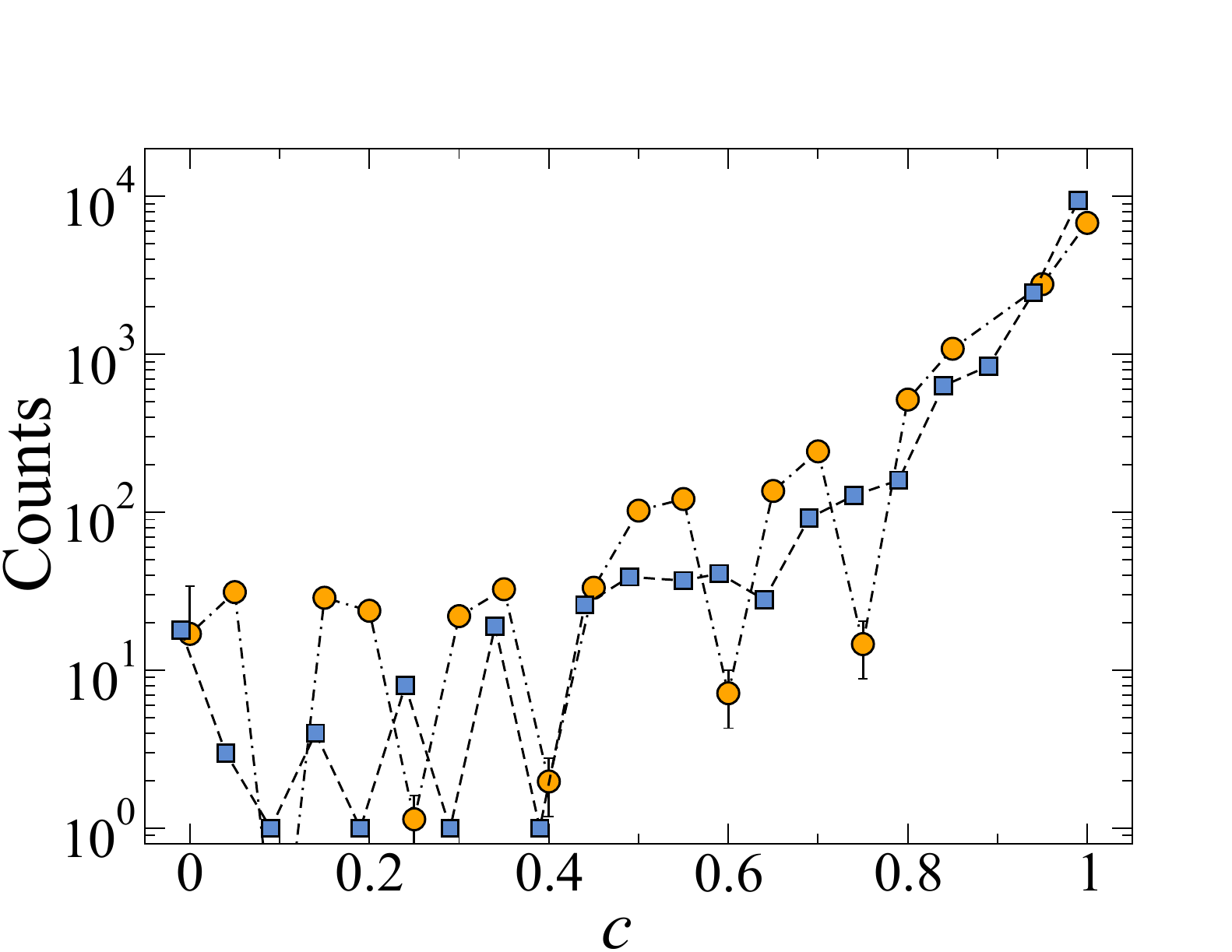}
\caption{Model results and validation for R\&D collaborations of firms: {\it (a)} Degree
    distribution of the R\&D network obtained from the agent-based model (blue line) and from the empirical network (circles).  {\it (b)}
    Comparison of the coreness distribution for the R\&D network obtained from the agent-based model (orange) and from the empirical network (blue). The results are averaged over 100 model realizations and the error bars (when visible) indicate standard errors.}
  \label{fig:GTS-4}
\end{figure}

To demonstrate that the agent-based model is also able to reproduce other empirical findings without overfitting, 
we plot in  Figure\ref{fig:GTS-4}(a) the degree distribution of the network ensembles obtained from the 100 realizations, alongside of the empirical degree distribution for the R\&D network, both for the final time. 
The excellent match of the two distributions should be noted. 
We further plot in  Figure\ref{fig:GTS-4}(b) the distribution of the coreness values both from the empirical data and from the computer simulations of the R\&D network.
Here, we use the normalized coreness $C^{\prime}$ instead of $C_{F}$. 
While one could argue about some deviations between the two in the range of small coreness values, we note that the core-periphery structure of the network is well captured by the model -- without any additional assumptions.

\subsection{Improving network embeddedness: Chance or choice?}
\label{sec:impr-netw-embedd}

Our observations about the impact of network embedding on the partner selection raise the question whether this impact should be interpreted as a change in the \emph{strategy} of an agent in selecting its partners.

Such a change of strategy could indeed have a rational explanation, as follows. 
Agents new to the network may have little chances to get connected to core agents.
Therefore, in the absence of better alternatives, they may eventually team up with other newcomers or agents from the periphery with comparable coreness.
Together with their partners, they then try to improve their network position. 
However, at the time of maximum network integration, the competition with other agents of similar or lower coreness can become more important than the opportunity to further increase their (already optimal) position. 
So, while previous partners may have become competitors, successful agents more likely search for, and to team up with, newcomers with fresh ideas.

The question is whether the observed change in partner selection indeed follows a \emph{strategy}, i.e. a deliberative process  to become more successful, or whether the ``strategy'' is still the same but opportunities have changed.
Then, contrasting the above explanation, one could argue that differences in partner selection are caused by different opportunities to be involved in a collaboration.
To decide between these two alternative explanations, we can use 
our agent-based model because it allows to disentangle strategic behavior from probabilistic actions.
Precisely, in our model agents are assigned constant probabilities for linking to newcomers or incumbents. 
Hence, differences in observed actions are not expressed by the link probabilities, which are the same for all agents, but by the process to become initiators of collaborations, and to be selected within a partner category ``same/different/no label''.

Our results have demonstrated that this model is able to reproduce the observed change in partner selection together with other topological features, such as degree distribution and coreness distribution.
Therefore, we can conclude that the change in choosing partners can be reproduced \emph{without} assuming  changes in the selection rules.
This does \emph{not} allow to conclude that agents do not follow strategies in selecting their partners, or change these strategies dependent on the network position. 
But it demonstrates that agents do not need to \emph{change strategies} to act in a way that is observed in their evolution of network embeddedness. 

In fact, the abundance of collaboration opportunities is one of the driving forces behind the process of improving network embeddedness.
While it is true that agents with a high activity are more often selected, it is also true that the number of newcomers or peripheral agents is much larger than the number of core agents.
Whenever newcomers or peripheral agents are activated to become initiators, they follow the  preferential selection rule, i.e. they tend to select partner agents with a larger degree.
These are most likely agents from the core, which are well embedded in the network.
Because these agents are more often selected for collaboration, their degree increase over time which further increases their probability to be selected, next time.
A higher degree, on the other hand, relates to a lower coreness value as outlined above, albeit not in a linear manner. 
It is this feedback process that eventually helps some agents to further improve their network embeddedness, whereas the majority of agents still stays in the periphery.

\section{Discussion}
\label{sec:discussion}

In this paper we have focused on the dynamics of collaboration networks, using two data sets from different domains, about R\&D collaborations between firms and about co-authorship collaborations between scientists.
To answer our initial question, whether these different collaboration networks can be characterized and modeled from a unifying perspective, we made two contributions.

Firstly, we proposed a new measure for network embeddedness, the (relative) coreness value $c^{i}(t)$, to compare the topological positions of individual agents.
We'd like to emphasize that our measure of network embeddedness is also a good predictor of success of individual agents.
Looking at the R\&D collaboration network of firms and their corresponding patent data \citep{vaccario-verginer-2022-networ}, we verified that for the successful firms a better coreness comes along with more patents whereas for the ``normal'' firms both the position and the number of patents is rather level (in comparison to the successful firms).

Monitoring the change of individual coreness values, we found in both data sets the same dynamics (shown in Figure~\ref{fig:GTS-3}).
Some agents over time have improved their network embeddedness by moving from the \emph{periphery} of the collaboration network close to the \emph{core}, i.e. from high to low coreness values. 
From the data, we obtained a time $t^{i}_{c}$ for each agent when the minimal coreness, i.e. the best network embeddedness, is reached. 
At about $t_{c}^{i}$ we observed a \emph{change} in partner selection, from partners of \emph{similar} coreness to partners of \emph{different} (i.e. high) coreness.

To explain this seemlingly strategic behavior was one aim of our agent-based model, which is our second contribution.
We utilized a stochastic label propagation model with five probabilities to form collaboration links between newcomers and established agents with similar or different labels. 
These probabilities could be calibrated using aggregated quantities from the respective empirical collaboration networks.
As the result, we obtained a \emph{data-driven} agent-based model, where domain specific information is included in interpretable link probabilities. 
Additionally, we implemented a hypothesis of how agents choose collaboration partners \emph{within} the three partner categories same/different/no label, namely by preferentially choosing agents with a higher degree.

We could demonstrate that these modeling assumptions are sufficient to explain the observed impact of network embeddedness on the selection of collaboration partners.
In a nutshell, there exists a feedback between an agent's activity on the one hand and its ability to increase the degree by establishing new collaborations and to propagate the own label to newcomers, on the other hand. 
More collaborations not only lead to higher degrees, but also to lower coreness values, i.e. agents become embedded in the core of the network.
The chance to propagate the own label later increases the chance to be selected for new collaborations, because empirics has shown that firms or scientists prefer to collaborate with partners with the same label.

These combined effects eventually explain the observation that core agents tend to collaborate preferably with newcomers or agents from the periphery.  
In fact, newcomers and peripheral agents choose these core agents with larger probability, once they managed to establish their strong network embeddedness.
Thus, in conclusion, what seems to be a deliberative strategy of successful agents, namely to switch their rules of partner selection, can be basically explained without strategic considerations. 
These cannot be excluded, but the model suggests that the empirical observations do not already imply such considerations. 
Hence, the emergence of realistic core-periphery structures in collaboration networks can be successfully modeled  without deliberative agents.

\section*{Acknowledgments} 

The authors acknowledge financial support from the EU-FET project
MULTIPLEX 317532. The calculations of network measures, and the
statistical treatment of data were performed using the R software for
statistical analysis v2.15.2, and the igraph library v0.6.2.

\section*{Author Contributions}

A.G. and F.S. designed the research, A.G., M.V.T., GV and LV performed the research, and analyzed the data. F.S. and A.G. wrote the manuscript.

\small \setlength{\bibsep}{1pt}
\bibliographystyle{sg-bibstyle-nourl}
\bibliography{GTS-RND_CareerPaths}

\end{document}